\documentclass[twocolumn]{revtex4-1}
\usepackage{amsmath}
\usepackage{amsfonts}
\usepackage{amssymb}
\usepackage{graphicx}
\usepackage{booktabs}
\newcommand{\ra}[1]{\renewcommand{\arraystretch}{#1}}

\ifx\pdfoutput\@undefined\usepackage[usenames,dvips]{xcolor}
\else\usepackage[usenames,dvipsnames]{xcolor}
\IfFileExists{pdfcolmk.sty}{\usepackage{pdfcolmk}}{}
\fi
\usepackage[plainpages=false,pdfpagelabels,pagebackref=false,naturalnames=true,hyperindex=true,pdftitle={Graph Automorphism and Topological Characterization of Synthetic and Natural Complex Networks by Information Content},pdfauthor={Hector Zenil}]{hyperref}
\hypersetup{colorlinks=true,
urlcolor=Cerulean,linkcolor=BrickRed,citecolor=RoyalBlue,a4paper,
 pdfpagemode=None,
 pdfstartview=fitH}
\usepackage[all]{hypcap}

\begin{document}
\title{Correlation of Automorphism Group Size and Topological Properties with Program-size Complexity Evaluations of Graphs and Complex Networks}

\author{Hector Zenil$^{1,3}$}\email{hectorz@labores.eu}
\author{Fernando Soler-Toscano$^{2,3}$}
\author{Kamaludin Dingle$^{4,5}$}
\author{Ard A. Louis$^{4}$\\$^{1}$\small Unit of Computational Medicine, Karolinska Institute, Sweden.\\$^{2}$\small Grupo de L\'ogica, Lenguaje e Informaci\'on, Universidad de Sevilla, Spain.\\$^{3}$\small Algorithmic Nature Group, LABoRES, Paris, France.\\$^{4}$\small Rudolf Peierls Centre for Theoretical Physics, University of Oxford, UK. \\$^{5}$\small The Systems Biology DTC, University of Oxford, UK.}

\date{\today}

                                             
\begin{abstract}
We show that numerical approximations of Kolmogorov complexity ($K$) of graphs and networks capture some group-theoretic and topological properties of empirical networks, ranging from metabolic to social networks, and of small synthetic networks that we have produced. That $K$ and the size of the group of automorphisms of a graph are correlated opens up interesting connections to problems in computational geometry, and thus connects several measures and concepts from complexity science. We derive these results via two different Kolmogorov complexity approximation methods applied to the adjacency matrices of the graphs and networks. The methods used are the traditional lossless compression approach to Kolmogorov complexity, and a normalised version of a Block Decomposition Method (BDM) based on algorithmic probability theory.\\

\end{abstract}

\maketitle

\section{Introduction}
Graphs are an important tool for mathematically analysing many systems, from interactions of chemical agents, to ecological networks, to representing data objects in computer science~\cite{newman2010networks,newman2011structure}. An interesting perspective regarding such graphs is to investigate the complexity~\cite{bonchev,kim2008complex} or information content of a graph~\cite{mowshowitz2012entropy}. While Shannon information theory~\cite{adami2011information,mowshowitz2012entropy,mowshowitz2009entropy} and counting symmetries~\cite{xiao2008emergence,xiao2008network} have been applied to measure information content/complexity of graphs, little has been done, by contrast, to demonstrate the utility of Kolmogorov complexity as a numerical tool for graph and real-world network investigations. Some theoretical connections between graphs and algorithmic randomness have been explored (e.g.~\cite{buhrman}), but these are mostly related to formal properties of random graphs. 

Here we computationally study some of these numerical and real-world directions and show how Kolmogorov complexity can capture group-theoretic and topological properties of abstract and empirical graphs and networks. We do this by introducing a measure of graph complexity based on approximating the Kolmogorov complexity of the adjacency matrix representation of a graph, which we achieve by applying our recently developed \emph{Block Decomposition Method} (below)~\cite{kolmo2d}. A theoretical advantage of using Kolmogorov complexity $K$ is that the measure $K$ is designed to capture \emph{all} structure (i.e. non-randomness) in an object, such as a graph.  In contrast, only looking at symmetries, for example, can miss structure and potential simplicity in a graph. For example, Fig.~\ref{fig:simplegraph} shows a graph with no symmetries, despite being far from random, and indeed is intuitively simple.

This paper is structured as follows: The next two Sections give background definitions and theorems; then in Section IV we consider the question of whether approximations of Kolmogorov complexity can characterise group-theoretic and topological properties of graphs and networks, more specifically, we study the number of automorphisms of a graph and introduce a normalised measure of graph complexity; following this in Section V we apply our measure to real-world networks; and finally we study in Section VI algorithmic information-theoretic similarities of networks with similar topologies created using different mechanisms (e.g. random versus preferential attachment). 


\section{Preliminaries}
\label{preliminaries}

\subsection{Graph notation}

A graph $g=(V,E)$ consists of a set of vertices $V$ (also called nodes) and a set of edges $E$. Two vertices, $i$ and $j$, form an edge of the graph if $(i,j)\in E$. Let the binary adjacency matrix of $g$ be denoted by $Adj(A)$. A graph can be represented by its adjacency matrix. Assuming that the vertices are indices from 1 to $n$, that is, that $V=\{1, 2, \ldots, n\}$, then the adjacency matrix of $g$ is an $n\times n$ matrix, with entries $a_{i,j}=1$ if $(i,j)\in E$ and 0 otherwise. The distance $D(g)$ of a graph $g$ is the maximum distance between any 2 nodes of $g$. The size $V(g)$ of a graph $g$ is the vertex count of $g$; similarly $E(g)$ denotes the edge count of $g$.\\

\noindent \textbf{Definition 1.} Two graphs $g$ and $h$ are \emph{isomorphic} if and only if there exists a permutation $\lambda$ such that $\lambda(g) = h$. (That is, $g$ and $h$ are topologically equivalent).\\

\begin{figure}[htbp!]
  \includegraphics[width=5.0cm]{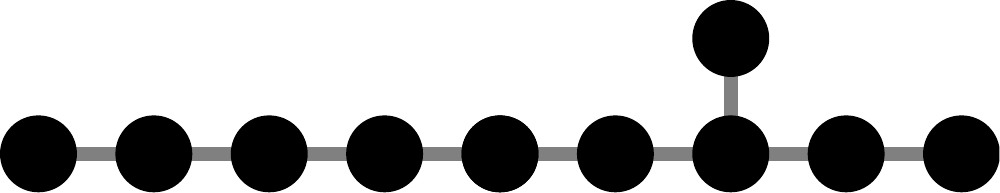}
  \caption{\label{fig:simplegraph} An example of a graph which has no symmetries (i.e. automorphism group size of 1, cf. Def. 2), despite being intuitively simple, i.e. just a string of nodes with a single side node.}
\end{figure}

The general problem of graph isomorphism appears, for example, in chemistry~\cite{Hopcroft1974,read77} and biology~\cite{bask06,bask07}. \\ 

\noindent \textbf{Definition 2.} An \emph{automorphism} of a graph $g$ is a permutation $\lambda$ of the vertex set $V$, such that the pair of vertices $(i,j)$ forms an edge if and only if the pair $(\lambda(i),\lambda(j))$ also forms an edge.\\

The set of all automorphisms of an object forms a group, called the \emph{automorphism group}. Intuitively, the size of the automorphism group $A(g)$ provides a direct measure of the abundance of symmetries in a graph or network. Every graph has a trivial symmetry (the identity) that maps each vertex to itself.\\


A clustering coefficient is a measure of the degree to which nodes in a graph tend to cluster together (for example, friends in social networks~\cite{GirvanNewman}). 
\\

\noindent \textbf{Definition 3.} 
$$C(v_i)=\frac{2\left|E(N_i)\right|}{n_i (n_i-1)}$$ where $E(N_i)$ denotes the set of edges with both nodes in $N_i$, and  $C(v_i)$ is the local clustering coefficient of node $v_i$.  

\section{Approximating Kolmogorov complexity}
\label{kolmo}

\subsection{Background Results}
Before describing our graph complexity measure, we provide some pertinent definitions and results from the theory of algorithmic randomness.

The \emph{Kolmogorov complexity} of a string $s$ is the length of the shortest program $p$ that outputs the string $s$, when run on a universal Turing machine $U$. Formally \cite{kolmo,chaitin},\\

\noindent \textbf{Definition 4.}
\begin{equation}
K_U(s) = \min \{|p|, U(p)=s\}
\end{equation}

By the \emph{invariance theorem}~\cite{li}, $K_U$ only depends on $U$ up to a constant $c$, so as is conventional, the subscript $U$ of $K$ is dropped. A technical inconvenience of $K$ as a function taking $s$ to the length of the shortest program that produces $s$, is that $K$ is semi-computable. Lossless compression algorithms are, however, traditionally used to approximate $K$ given that compressibility is a sufficient test of low Kolmogorov complexity.

The concept of algorithmic probability (also known as Levin's semi-measure) yields a method to approximate Kolmogorov complexity. The algorithmic probability of a string $s$ is a measure that describes the probability that a random program $p$ produces the string $s$ when run on a universal (prefix-free \footnote{The group of valid programs forms a prefix-free set (no element is a prefix of any other, a property necessary to keep $0 < m(s) < 1$.) For details see~\cite{calude}.}) Turing machine $U$. Formally~\cite{solomonoff,levin,chaitin},\\

\noindent \textbf{Definition 5.} 
\begin{equation}
\label{codingeq}
m(s) = \sum_{p:U(p) = s} 1/2^{|p|}
\end{equation}

The probability semi-measure $m(s)$ is related to Kolmogorov complexity $K(s)$ in that $m(s)$ is at least the maximum term in the summation of programs ($m(s)\geq2^{-K(s)}$), given that the shortest program carries the greatest weight in the sum. The algorithmic Coding Theorem~\cite{cover,calude} further establishes the connection between $m(s)$ and $K(s)$.\\

\noindent \textbf{Theorem 1.} (\cite{levin}):
\begin{equation}\label{cdthm}
|-\log_2 m(s) - K(s)| < c
\end{equation}
where $c$ is some fixed constant, independent of $s$. The theorem implies~\cite{cover} (pp. 190-191) and~\cite{zenil2007,delahayezenil} that one can estimate the Kolmogorov complexity of a string from the frequency of production from running random programs by simply rewriting Eq.~(\ref{cdthm}) as:

\begin{equation}
K(s)=-\log_2 m(s) + O(1)  
\end{equation}

The advantage of calculating $m(s)$ as an approximation of $K$ by application of the Coding Theorem is that $m(s)$ retrieves finer-grained values than compression algorithms, which  are unable to compress and therefore distinguish complexity variations for small objects. Applications of $m(s)$ have been explored in~\cite{thesis,zenil2007,d4,zenilalgo,d5,numerical}, and include applications to image classification~\cite{kolmo2d} and to the evolution of space-time diagrams of cellular automata~\cite{zenilchaos}. Error estimations of approximations of $m(s)$ have also been reported before in~\cite{kolmo2d}, showing that estimations of the constant involved in the invariance theorem remain small when taking larger samples of
Turing machines from which $m(s)$ is calculated. Moreover, in the same paper~\cite{kolmo2d}, it is shown that values between $m(s)$ and lossless compression algorithms, for experiments where both methods overlap, remain close to each other. Here Fig.~\ref{dual} provides an indication of the error estimation as compared to the estimation error of the lossless compression algorithm used (Deflate), by comparing the Kolmogorov complexity of dual graphs estimated using both $m(s)$ and lossless compression.

\subsection{The Coding Theorem Method}
In~\cite{d4} a technique was advanced for approximating $m(s)$ (hence $K$) by means of a function that considers all Turing machines of increasing size (by number of states). Let $(n,k)$ denote the set of Turing machines with $n$ states and $k$ symbols using the Busy Beaver formalism~\cite{rado} and let $T$ be a Turing machine in $(n, k)$ with empty input. Then:\\

\noindent \textbf{Definition 6.} 
\begin{equation}
\label{Deq} 
\mathbb{D}(n, k)(s)=\frac{|\{T\in(n, k) : T \textit{ produces } s\}|}{|\{T \in(n, k) : T \textit{ halts }\}|}
\end{equation}

For small values $n$ and $k$, $\mathbb{D}(n, k)$ is computable for values of the Busy Beaver problem that are known. The Busy Beaver problem is the problem of finding the $n$-state, $k$-symbol (or colour) Turing machine which writes a maximum number of non-blank symbols on the tape before halting and starting from an empty tape, as originally proposed by Rado~\cite{rado}. Or alternatively, finding the Turing machine that performs a maximum number of steps when started on an initially blank tape before halting. For $n=4$ and $k=2$, for example, the Busy Beaver machine has maximum runtime $S(n)=107$~\cite{brady}, from which one can deduce that if a Turing machine with 4 states and 2 symbols running on a blank tape hasn't halted after 107 steps then it will never stop. This is how $\mathbb{D}$ was initially calculated with the help of the Busy Beaver. However, the Busy Beaver problem is only computable for small $n,k$ values. Nevertheless, one can continue approximating $\mathbb{D}$ for a greater number of states (and colours) proceeding by sampling as described in~\cite{d5}, with an informed runtime after studying the behaviour of the runtime distribution of the Busy Beavers.

The Coding Theorem Method~\cite{d4,d5} is rooted in the relation provided by algorithmic probability between frequency of production of a string from a random program and its Kolmogorov complexity (Eq.~(\ref{cdthm})). Essentially it uses the fact that the more frequent a string is, the lower Kolmogorov complexity it has; and strings of lower frequency have higher Kolmogorov complexity.

\subsection{Applying the Block Decomposition Method to graphs}
\label{sec:two-dimens-turing}

As an extension of the Coding Theorem Method~\cite{d4,d5} of approximating bit string complexity (above), a method to approximate the Kolmogorov complexity of $d$-dimensional objects was advanced in~\cite{kolmo2d}. The method is called the \emph{Block Decomposition Method} (BDM) and it consists of decomposing larger objects into smaller pieces for which complexity values have been estimated, then reconstructing an approximation of the Kolmogorov complexity of the larger object by adding the complexity of the individual pieces according to rules of information theory.

By taking the adjacency matrix of a graph as the representation of the graph, we can use the BDM to estimate $K$ for graphs. Given the 2-dimensional nature of adjacency matrices we can use a variation of a Turing machine that runs on a 2-dimensional tape in order to estimate upper bounds of $K$ of the adjacency matrix of a graph. A popular example of a 2-dimensional tape Turing machine is Langton's ant~\cite{langton}. Another way to see this approach is to take the BDM as a form of deploying all possible 2-dimensional deterministic Turing machines of a small size in order to reconstruct the adjacency matrix of a graph from scratch (or smaller pieces that fully reconstruct it). Then as with the Coding Theorem Method (above), the Kolmogorov complexity of the adjacency matrix of the graph can be estimated via the frequency that it is produced from running random programs on the (prefix-free) 2-dimensional Turing machine. 

Having outlined the theory, here we will use Eq.~(\ref{Deq}) with $n=5$ and $k=2$ to approximate $K$. The choice of this $n$ and $k$ is because 2-dimensional Turing machine with $n=4$ states and empty input produce all square arrays (or matrices) of size 3 by 3 but not all of size 4 by 4. The larger the arrays, the better approximations of $K$ for large objects decomposed in square arrays, and $n=5$ was a Turing machine size that we were able to run with current technology on a medium-size supercomputer ($\sim 30$ cpus) for about a month. Details are provided in~\cite{d5}. One can continue calculating further and approximations will improve accordingly.

Let $\mathbb{D}(5,2)$ be the frequency distribution constructed from running 2-dimensional machines (hence producing arrays rather than strings) according to Eq.~(\ref{Deq}). Then, for an array $s$,\\

\noindent \textbf{Definition 7.} 
\begin{equation}
\label{ecaeq}
K_m(s) = -\log_2(\mathbb{D}(5,2)(s))
\end{equation}
where $K_m(s)$ is an approximation to $K$ by means of the Coding Theorem. $K_m(s)$ will then be used to estimate the complexity of graphs as follows: Formally, we will say that a graph $g$ has complexity:\\

\noindent \textbf{Definition 8.} 
\begin{equation}
\label{newecaeq}
K\log_m(g) = \sum_{(r_u,n_u)\in Adj(g)_{d\times d}} \log_2(n_u)+K_m(r_u)
\end{equation}
where $Adj(g)_{d\times d}$ represents the set with elements $(r_u,n_u)$, obtained when decomposing the adjacency matrix of $g$ into non-overlapping squares of size $d$ by $d$. In each $(r_u,n_u)$ pair, $r_u$ is one such square and $n_u$ its multiplicity (number of occurrences). From now on both $K_m$ and $K\log_m$ will denote the same Eq.~(\ref{newecaeq}) given that Eq.~(\ref{ecaeq}) will no longer be used.

In~\cite{kolmo2d} a set of 2-dimensional Turing machines was executed to produce all square arrays of size $d=4$. This is why BDM is needed in order to decompose objects of larger size into objects for which its Kolmogorov complexity has been estimated. Using squares of size 4 is like looking for local regularities that give an approximation of the Kolmogorov complexity of the larger object. Because we will take $d=4$ from this point on we will denote by $K_m$ (sometimes also by $K\log_m$, but $m$ should not be taken as the base of the $\log$, which is 2 in the definition of $K\log_m$). The reader should simply bear in mind that $K_m$ and $K\log_m$ are approximations of Kolmogorov complexity using the Block Decomposition Method (BDM).

\subsection{Complexity and graph vertex order}
\label{sec:compl-order-graph}

\begin{figure}[htbp!]
  \centering
  \includegraphics[width=4.5cm]{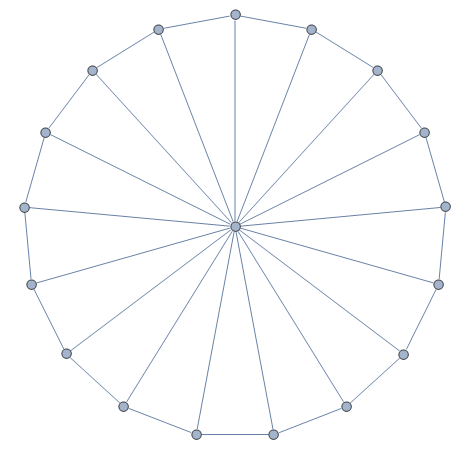}
  \caption{Wheel-18 graph.}
  \label{fig:wheel18a}
\end{figure}

\begin{figure}[htbp!]
  \centering
  \includegraphics[width=8.5cm]{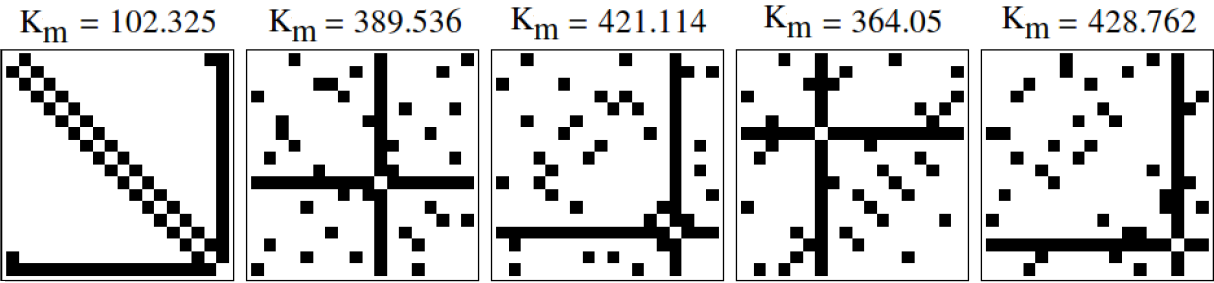}
  \caption{Wheel-18 graph with the nodes in five different orders. Despite each adjacency matrix corresponding to the same topology (i.e. the Wheel-18 graphs), the different node orders affect the estimated complexity. }
  \label{fig:wheel18}
\end{figure}

As a final note on our method of approximating $K$, the order of the graph nodes in the adjacency matrix is relevant for the complexity retrieved by BDM. This is especially important in highly symmetrical graphs. Fig.~\ref{fig:wheel18} shows five different adjacency matrices for the wheel graph of 18 nodes (Fig.~\ref{fig:wheel18a}), with the nodes ordered in different ways. The extreme left matrix of Fig.~\ref{fig:wheel18} represents the nodes in consecutive order, with the central one at the right. The other four matrices sort the nodes in random order. As we can see in this very regular graph, the lowest complexity corresponds to the organized matrix.

Hence, when studying the complexity of graphs, for several applications we will be interested not only in the $K_m$ value of a particular adjacency matrix, but in several randomisations of the vertex order. Notice that it is just the adjacency matrix representation that changes; topologically the graph is the same in all permutations of the nodes. In estimating complexity, it is reasonable to consider that the complexity of a graph corresponds to the lowest $K_m$ value of all permutations of the adjacency matrix, as the shortest program generating the simplest adjacency matrix is the shortest program generating the graph. Hence when estimating $K_m$ we find a large number of programs producing the array from which a minimum complexity value is calculated by means of the frequency of the programs producing the array.

\subsection{Normalising BDM}

We now also introduce a normalised version of the BDM. The chief advantage of a normalised measure is that it enables a comparison among objects of different sizes without allowing the size to dominate the measure. This will be useful to compare graphs and networks of different sizes. First, for a square array of size $n\times n$, we define:
\[
MinBDM(n)_{d\times d} = \lfloor n/d \rfloor + \displaystyle\min_{x\in M_d(\{0,1\})} K_m(x)
\]
\noindent Where $M_d(\{0,1\})$ is the set of binary matrices of size $d\times d$. For any $n$, $MinBDM(n)_{d\times d}$ returns the minimum value of Eq.~\eqref{newecaeq} for square matrices of size $n$, so it is the minimum BDM value for graphs with $n$ nodes. It corresponds to an adjacency matrix composed by repetitions of the least complex $d\times d$ square. It is the empty (or complete) graph, because $0_{d,d}$ and $1_{d,d}$ are the least complex squares (the most compressible) of size $d$.

Secondly, for the maximum complexity, Eq.~(\eqref{newecaeq}) returns the highest value when the result of dividing the adjacency matrix into $d\times d$ squares contains the highest possible number of different squares (to increase the sum of the right terms in Eq.~\eqref{newecaeq}) and the repetitions (if necessary) are homogeneously distributed along those squares (to increase the sum of the left terms in Eq.~\eqref{newecaeq}), which should be the most complex ones in $M_d(\{0,1\})$. For $n,d\in \mathbb{N}$, we define a function \[f_{n,d}: M_d(\{0,1\})\longmapsto \mathbb{N}\]
that verifies:
\begin{eqnarray}
  \label{eq:2}
  & \displaystyle\sum_{r\in M_d(\{0,1\})} f_{n,d}(r) = \lfloor
  n/d\rfloor^2 & \\
  \label{eq:3}
  & \displaystyle\max_{r\in M_d(\{0,1\})} f_{n,d}(r)\ \ \leq\ \ 1+
  \displaystyle\min_{r\in M_d(\{0,1\})} f_{n,d}(r)  & \\
  \label{eq:4}
  & 
  K_m(r_i) > K_m(r_j)\ \Rightarrow \ f_{n,d}(r_i)\geq
  f_{n,d}(r_j)
  &
\end{eqnarray}

The value $f_{n,d}(r)$ indicates the number of occurrences of $r\in
M_d(\{0,1\})$ in the decomposition into $d\times d$ squares of the
most complex square array of size $n\times n$. Condition Eq.~\eqref{eq:2}
establishes that the total number of component squares is $\lfloor
n/d\rfloor^2$. Condition Eq.~\eqref{eq:3} reduces the square repetitions
as much as possible, to increase the number of differently
composed squares as far as possible and distribute them
homogeneously. Finally, Eq.~\eqref{eq:4} ensures that the most complex
squares are the best represented. Then, we define:

\[
MaxBDM(n)_{d\times d} = \hspace{-0.5cm}\sum_{
  {\begin{array}{c}
    r\in M_d(\{0,1\}),\\f_{n,d}(r)>0
  \end{array}}} \hspace{-0.5cm}
\log_2(f_{n,d}(r))+ K_m(r)
\]

Finally, the normalised BDM value of a graph $g$ is:\\

\noindent \textbf{Definition 9.} Given graph $g$ with $n$ nodes,
$NBDM(g)_d$ is defined as 
\begin{equation}
\label{nbdm}
\frac{K_m(g) - MinBDM(n)_{d\times
    d}}{MaxBDM(n)_{d\times d} - MinBDM(n)_{d\times d}}
\end{equation}

This way we take the complexity of a graph $g$ to have a normalised value which is not dependent on the size $V(g)$ of the graph but rather on the relative complexity of $g$ with respect to other graphs of the same size. The use of $MinBDM(n)_{d\times d}$ in the normalisation is relevant. Note that the growth of $MinBDM(n)_{d\times d}$ is linear with $n$, and the growth of $MaxBDM(n)_{d\times d}$ exponential. This makes that for complex graphs, the result of normalising using just $K_m(g)/MaxBDM(n)_{d\times d}$ would be similar to $NBDM(g)_d$. But it would not work for simple graphs, as when the complexity of $g$ is close to the minimum, the value of $K_m(g)/MaxBDM(n)_{d\times d}$ drops exponentially with $n$. For example, the normalised complexity of an empty graph would drop exponentially in its size. To avoid it, Eq.~\eqref{nbdm} considers not only the maximum but also the minimum.

Notice the heuristic character of $f_{n,d}$. It is designed to ensure a quick computation of $MaxBDM(n)_{d\times d}$, and the distribution of complexities of squares of size $d\in\{3,4\}$ in $\mathbb{D}(5,2)$ ensures that $MaxBDM(n)_{d\times d}$ is actually the maximum complexity of a square matrix of size $n$, but for other distributions it could work in a different way. For example, condition~\eqref{eq:3} assumes that the complexities of the elements in $M_d(\{0,1\})$ are similar. This is the case for $d\in\{3,4\}$ in $\mathbb{D}(5,2)$, but
it may not be true for other distributions. But at any rate it offers a way of comparing the complexities of different graphs independent of their size.\\

Having defined our complexity measure and its normalised version, we verify it by showing that it behaves in accordance with theory when analysing edge density and dual graphs.

\subsection{Complexity and edge density}
\label{sec:compl-numb-edges}

\begin{figure}[htbp!]
  \centering
  \includegraphics[width=6.6cm]{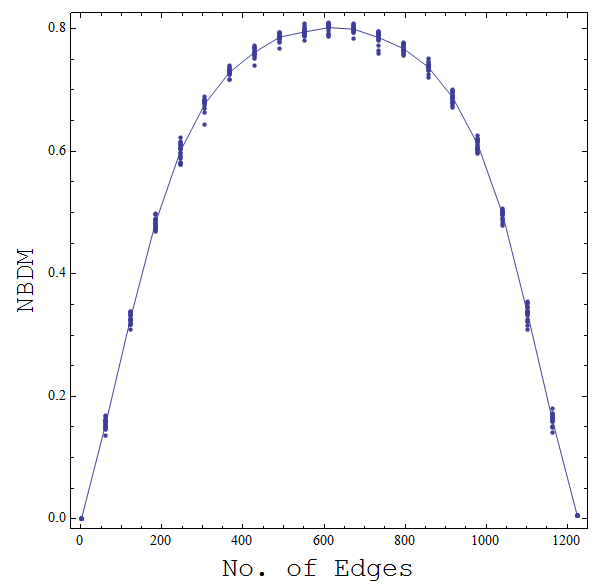}
  \includegraphics[width=6.6cm]{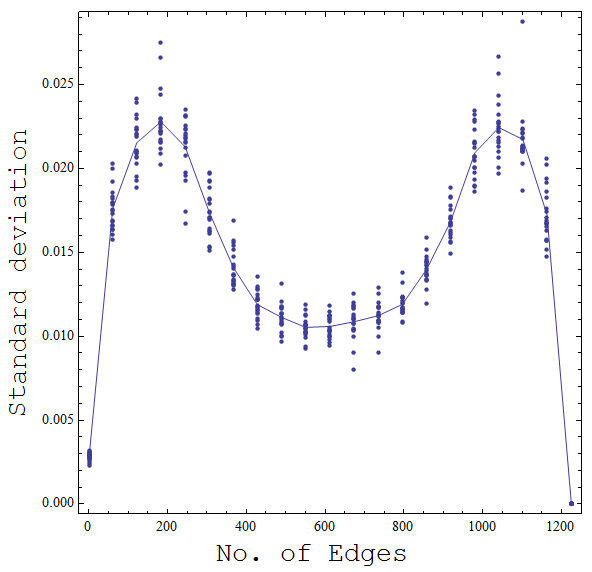}
  \caption{Estimated (normalised) Kolmogorov complexity for increasing number of edges for random graphs of 50 nodes each. The minimum complexity (Top) and standard deviation (Bottom) is shown for 100 random permutations of 20 graphs in each group.}
  \label{fig:50meanStdDev}
\end{figure}

There has been a protracted discussion in the literature as to whether the complexity of a structure increases with its connectivity, beginning with a disconnected graph with no edges, or whether instead it reaches a maximum before returning to zero for complete graphs~\cite{bonchev}. In~\cite{gellmann}, for example, Gell-Mann asks about the algorithmic complexity (description length) of small graphs with eight vertices each and an increasing number of edges $E(g)=0$ to $E(g)=V(g)(V(g)-1)/2$ (complete graph). Gell-Mann reasonably argues that these two extreme cases should have roughly equal complexity. Graphs with 0.5 edge density should fall in between and the other cases are more difficult to tell apart. Here we provide an answer to the general question of the relation between description complexity and edge count, an answer which is in agreement with Gell-Mann while being at odds with several other tailor made measures of complexity~\cite{bonchev}, \cite{dehmer1}, \cite{dehmer2} and \cite{standish}. We have created a number of random graphs, all with 50 nodes but different numbers of edges, ranging from 1 to ${50 \choose 2} = 1225$, in intervals of $1225/20$ edges. For each interval, we created 20 random graphs and generated 100 random permutations of each (see Fig.~\ref{fig:50meanStdDev}). All the graphs (Fig.~\ref{fig:50meanStdDev}) in the same interval correspond to aligned points. There are always 20 aligned points, but in most cases they overlap. Points represent the minimum (Top) and standard deviations (Bottom) of the complexity of the 100 permutations of each group.

In fact, that the most complex binary strings (hence binary adjacency matrices) will have roughly equal zeros and ones is a standard result from Kolmogorov complexity theory. This can be shown straightforwardly by the following relation of $K$ to Shannon entropy: If $x=x_1,\dots,x_l$ is a bit string, then~\cite{cover},
 
\begin{equation}\label{eq:Shan}
K(x)\leq l H_0\left(\frac{1}{l}\sum_{i=1}^{l}x_i\right)+O(\log_2(l))
\end{equation}
where 
\begin{equation} 
H_0(p)=-p\log_2(p)-(1-p)\log_2(1-p)
\end{equation}
 is the Shannon entropy (in bits) of the string $x$. This then implies that if the number of edges diverges from roughly half of the maximum possible, then $p$ must diverge from 0.5, and so the entropy $H_0$ decreases, implying that $x$ cannot be algorithmically random. Hence, the most complex strings must have roughly equal zeros and ones, or in graph terms, they must have roughly half the number of possible edges. Additionally, Eq.~(\ref{eq:Shan}) predicts that Fig. \ref{fig:50meanStdDev} (Top) should look roughly like the graph of $H_0(p)$ vs $p$, which it does: peaking at the centre, with gradual decay to zero at $p=0,1$. Also, note the clear symmetry of the plots in Fig.~\ref{fig:50meanStdDev}. This is because the value of $K_m$ is the same for any given graph and its complement. Complexity is minimal for empty or complete graphs (the most homogeneous matrices, all 0 or all 1), and so the standard deviation is also minimal. 
These observations show that our measure is behaving as expected from theory. 

\subsection{Graph duality}
\label{sec:duality}

The dual graph of a planar graph $g$ is a graph that has a vertex corresponding to each face of $g$, and an edge joining two neighbouring faces for each edge in $g$. If $g\prime$ is a dual graph of $g$, then $A(g\prime)=A(g)$, making the calculation of the Kolmogorov complexity of graphs and their dual graphs interesting because of the correlation between Kolmogorov complexity and $A(g\prime)$, which should be the same for $A(g)$. One should also expect the estimated complexity values of graphs to be the same as those of their dual graphs, because the description length of the dual graph generating program is $O(1)$.

Note that unlike the t-statistic, the value of the D statistic (and hence the P value) is not affected by scale changes like using log. The KS-test is a robust test that cares only about the relative distribution of the data.

\begin{figure}[htbp!]
  \centering
\includegraphics[width=7.5cm]{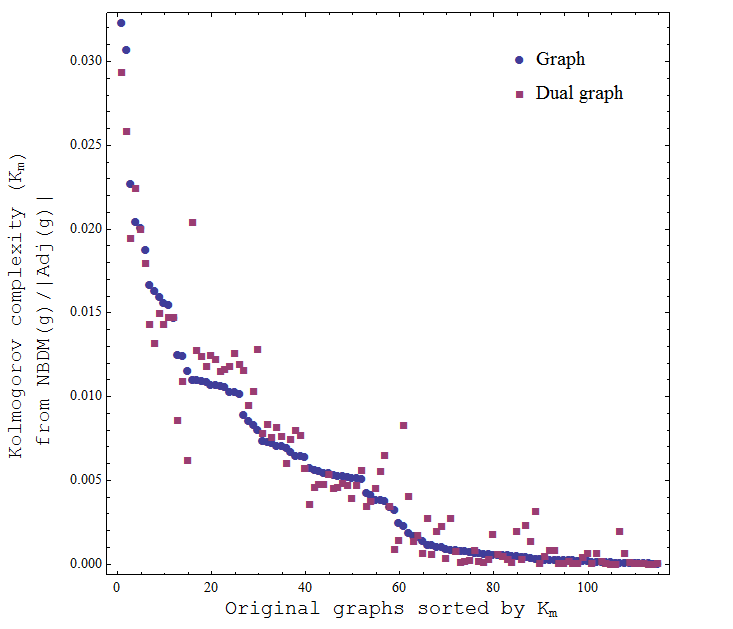}
\includegraphics[width=7.3cm]{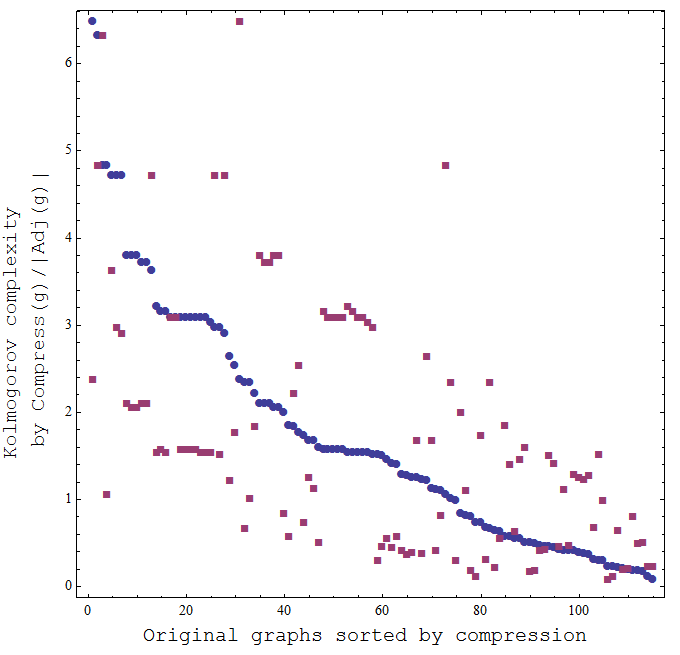}
  \caption{\label{dual} Log scatter-plots of graphs ranked by Kolmogorov complexity approximated by two different methods. Top: Dual graphs not ranked by the BDM method closely follow the distribution of their respective duals as one would expect from a complexity measure given that an $O(1)$ program can build a graph from its dual and vice-versa. Bottom: Lossless compression in agreement with BDM and the expected complexity of dual graphs.}
\end{figure}

We numerically approximated the Kolmogorov complexity of dual graphs
using both lossless compression and the BDM applied to the adjacency matrices
of 113 regular graphs with non-identical dual graphs found in
\emph{Mathematica}'s built-in repository GraphData[]. The values (see
Figs.~\ref{dual}) were normalised by a multiple of the size of the
adjacency matrices $c|Adj(A)|$. Graphs ($g$) and their dual graphs
($g\prime$) were found to have estimated Kolmogorov complexity values that
are close to each other. The Spearman coefficient $r$ between $K_m(g)$ and
$K_m(g\prime)$ NBDM estimations amounts to $r=0.96$.
Values approximated by lossless compression were calculated with
Deflate (a standard compression algorithm) implemented in
\emph{Mathematica}'s Compress[] function. Notice that the BDM (and
therefore the NBDM) accumulates errors for larger graphs if the
complementary algorithm (CTM) is not run for a greater number of
Turing machines. A more accurate analysis of divergence rates and
accumulated errors should be investigated. Robustness of algorithmic probability approximations for different (Turing-universal) computational formalisms (e.g. deterministic cellular automata, Post tag systems, etc.) was investigated in~\cite{zenilalgo}, where frequency distributions were calculated and their ranking order correlation also quantified with the Spearman coefficient.

These results show that Kolmogorov complexity applied to adjacency matrices of graphs as a measure of complexity behaves as expected, agreeing both with theory and intuition. Having verified that our measure behaves as expected, we can now apply our measure to analysing automorphism in graphs.

\section{Graph automorphisms and Kolmogorov complexity}

Intuition tells us that a graph with more symmetries can be more compactly described and should therefore have a lower Kolmogorov complexity. For example, if an object is symmetrical a single symbol can be used for each repetition, in addition to an additive constant of information describing the type of transformation (e.g. rotation, reversion, translation). Specifically, any collection of graph nodes within the same orbit of a group transformation will require only the bonds of one node to be specified, with the other nodes then using the same information to specify their bonds. Hence, the number of bits required to specify the graph would be signifcantly lowered, reducing the graph Kolmogorov complexity. Consequently, as the size of its automorphism group $A(g)$ of a graph $g$ measures the extent of symmetries in $g$, one would expect to find Kolmogorov complexity to be related to $A(g)$. 

\begin{figure}[htbp!]
  \centering
  \includegraphics[width=8.8cm]{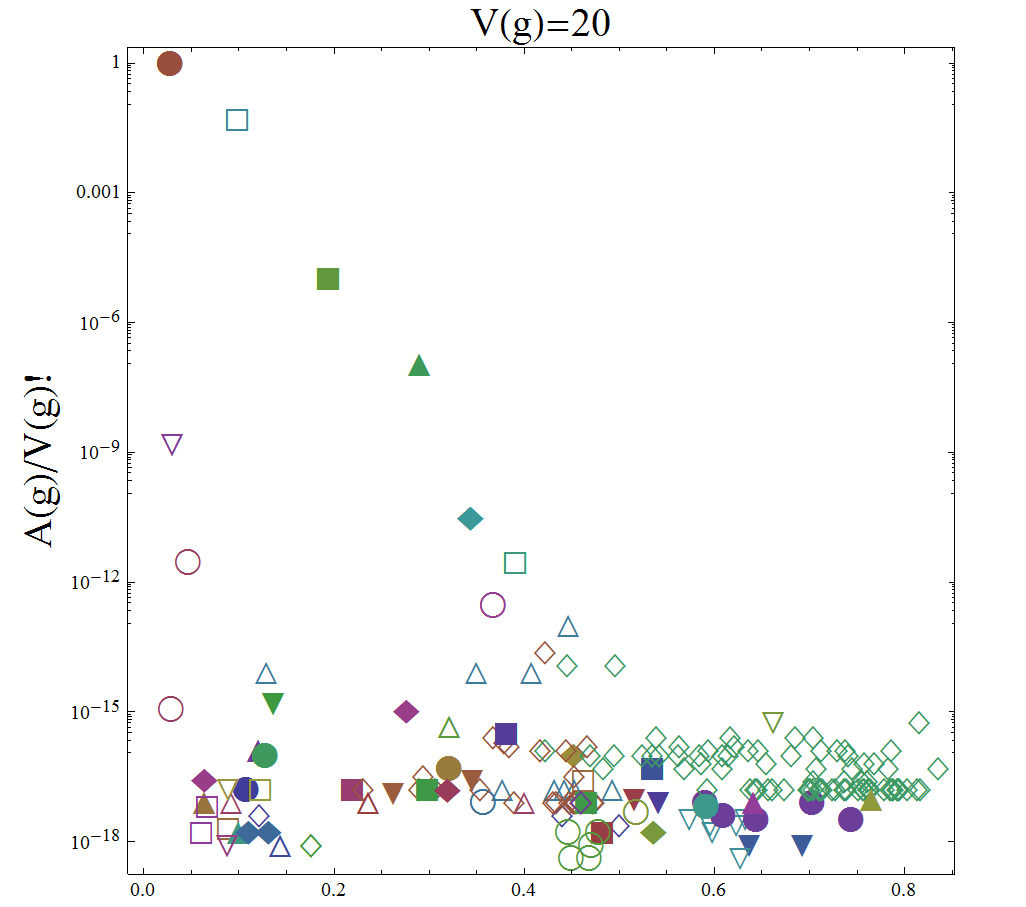}
\flushright
   \includegraphics[width=8.7cm]{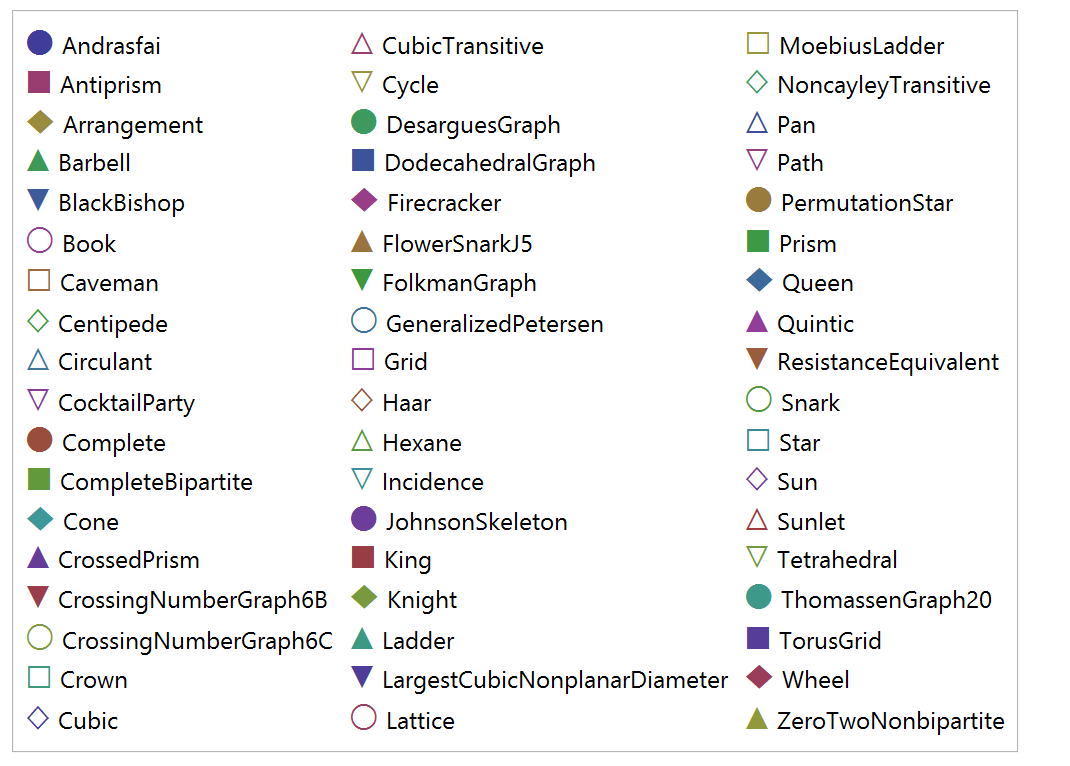} \caption{\label{fig7} Plot of graph automorphism group size $A(g)$ ($y$-axis) of all connected regular graphs of size $V(g)=20$ available in \emph{Mathematica}'s GraphData[] versus Kolmogorov complexity ($x$-axis) estimated by BDM. As theoretically expected, the larger automorphism group size $A(g)$ the smaller Kolmogorov complexity estimations.}
\end{figure}

We test this reasoning on connected regular graphs of size $V(g)=20$ nodes; Fig.~\ref{fig7} shows that $A(g)$ and $K_m(g)$ are indeed negatively related, as expected. 






\begin{figure}[htbp!]
  \centering
  \includegraphics[width=8.8cm]{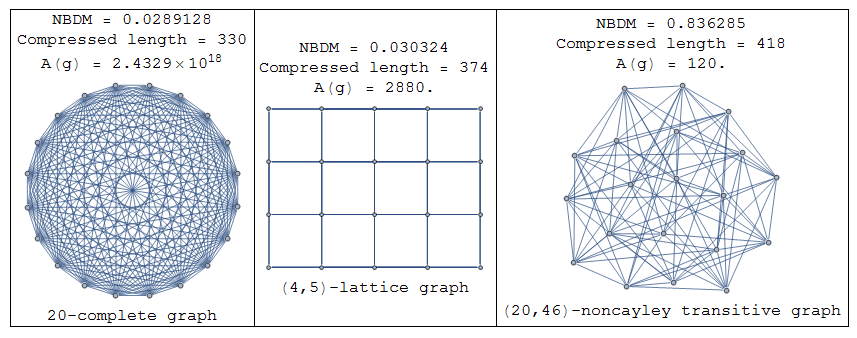}
  \caption{\label{fig7b} The three graphs found in the boundaries of Fig.~\ref{fig7}. From left to right: The graph at the top left, with low $K_m$ and large automorphism group size is the complete graph for $V(g)=20$. Bottom left with low $K_m$ but small automorphism group size is the (4,5)-lattice. Bottom right, with high $K_m$ and small automorphism group size: the (20,46)-noncayley transitive graph.}
\end{figure}

It is interesting that there are several graphs of low $K_m$ and also low $A(g)$ (see Fig.~\ref{fig7b})---that is, there are several graphs which have few symmetries but yet also low complexity (this is analogous to Fig.~\ref{fig:simplegraph}). Hence our measure is picking up structure in the graphs which are not detectable by symmetry search. 

We plotted analogous plots to Fig.~\ref{fig7} (i.e. connected regular graphs) for different number of nodes, using $V(g)$ between 22 and 36 (see Appendix Fig.~\ref{fig8} and Fig.~\ref{fig9}). The results were qualitatively the same, with graphs of larger $K$ estimations having smaller $A(g)$ values, and the results also agreed with those using lossless compression (Deflate) instead of BDM.

Notice that the measure $K_m$ does not quantitatively agree with the theoretical $K$, as the theoretical $K$ has an upper bound of $\sim$200 bits, which would be arrived at by specifying $Adj(g)$ literally and in full, i.e. using the fact that 
\begin{equation}
K(g)\leq {V(g) \choose 2}+2\log_2 {V(g) \choose 2} +O(1)
\end{equation}
 On the other hand, BDM retrieves values up to $\sim$800. Nonetheless, BDM values (just like compression results) are consistent upper bounds. BDM can provide better approximations but it requires 
 the calculation of a larger sample of random programs~\cite{d5} 
 with square matrices of larger size $d$, compared to the current $d=4$ that the experiments here introduced used. But in order for BDM to \emph{scale up} and provide better $K_m$ approximations to the theoretical (and ultimately uncomputable) $K$, the full method to consider is both Coding Theorem Method (CTM) $ + $ BDM. CTM is, however, computationally very expensive, while BDM is computationally very cheap, so there is a trade-off in the application of the algorithm. BDM alone is limited by the data generated by CTM, and the range of application of BDM will be limited in this respect, specially for increasingly larger objects (where lossless compression can take over and behave better than for small objects, hence CTM $ + $ BDM and lossless compression are complementary tools). The chief advantage of the CTM $ + $ BDM approach, is that CTM needs to run only once and BDM can be then efficiently applied and used many times on all sorts and types of data.

\section{Applying BDM to real-world natural and social networks}

\begin{figure}[htbp!]
  \centering
  \includegraphics[width=7.4cm]{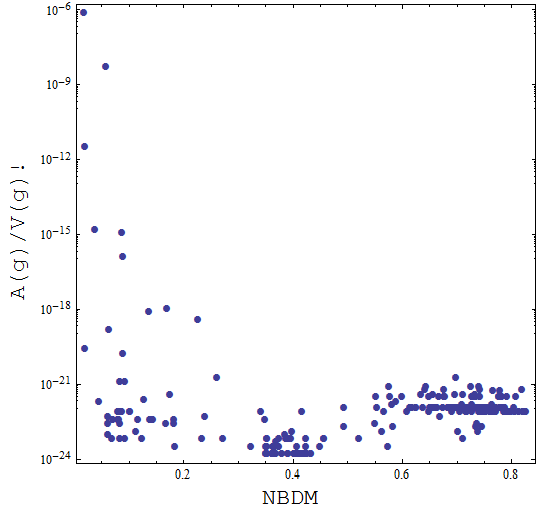}
  \caption{\label{largenetworks} Real-world networks also display the same correlation between Kolmogorov complexity and automorphism group size $A(g)$. Networks with more symmetries have lower estimated Kolmogorov complexity. Automorphisms count is normalised by network size.}
\end{figure}

We now move on to applying the normalised version of the BDM (NBDM) to real world networks of different sizes. 
The 88 real-world networks range from metabolic to social networks and have between 200 and 1000 nodes, and were extracted from the function ExampleData[``NetworkGraph"] in Wolfram \emph{Mathematica} v.9. A subset of 20 of the specific real-world network examples used in this experiment are in Table~\ref{table1} (Appendix).  Fig.~\ref{largenetworks} shows complexity plotted against automorphism group size for these networks. We find that the same qualitative relationship between $K$ and $A(g)$ as reported in Fig.~\ref{fig7} and Fig.~\ref{fig8} for synthetic networks (see Appendix) are obtained for these larger real-world networks. This provides further evidence that approximations of Kolmogorov complexity identify group-theoretic properties (as well as topological properties, as will be shown in Section~\ref{topossection}) of graphs and networks. The automorphism group sizes $A(g)$ were calculated using the software Saucy 3.0 (\url{http://vlsicad.eecs.umich.edu/BK/SAUCY/},  accessed in September 2013), the most scalable symmetry-finding tool available today~\cite{18}. Saucy only deals with undirected graphs hence only directed versions of the real-world sample of networks was used to calculate the automorphism group size $A(g)$ of each with Saucy. We made directed graphs into undirected graphs by setting an edge between two nodes $i$ and $j$ if in the original directed graph there was a directed edge from $i$ to $j$, $j$ to $i$, or both. Clearly undirected graphs are simpler than directed graphs, in general. In the case of labelled nodes, it is easy to see that the Kolmogorov complexity of a random undirected graph is typically half of a directed labelled graph, as the binary adjacency matrix for undirected graphs is symmetric. For unlabelled graphs, it is not so straightforward, due to complications of isomorphism (recall that there are typically many adjacency graphs representing a given unlabelled graph). Nonetheless directed graphs typically require more bits to specify their links, and are hence more complex. Exploring directed graphs will be left for future work.

\section{The algorithmic randomness of synthetic complex networks}
\label{topossection}

An objective and universal measure of complexity should take into account symmetries as a simplifying factor when it comes to description size. 
 Here we explore how $K$ can characterise topological properties of complex networks.

The study of complex networks is currently an active area of research~\cite{newman2011structure}. The field has been driven largely by observations that many real-world networks (e.g. internet links or metabolic networks) have properties very different from both regular and random graphs; the latter having been extensively studied in foundational work by Paul Erd\"os and Alfr\'ed R\'enyi. Specifically, two topological properties of many complex networks that have been a focus of interest are (a) a scale-free (or power law) distribution in node degree distributions, and (b) the ``small-world" property where graphs have high clustering and the average graph distance $D$ grows no faster than the $\log$ of the number of nodes: $D \sim \log(V(g))$. 

\begin{figure}[htbp!]
  \centering
  \includegraphics[width=2.7cm]{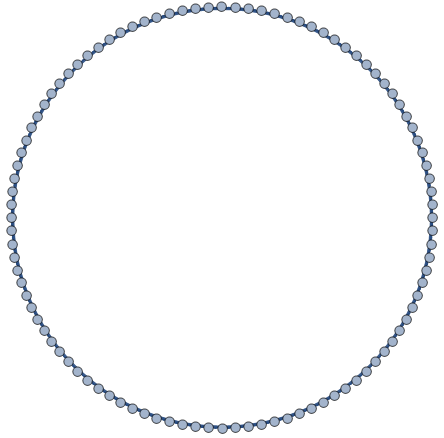}
 \includegraphics[width=2.7cm]{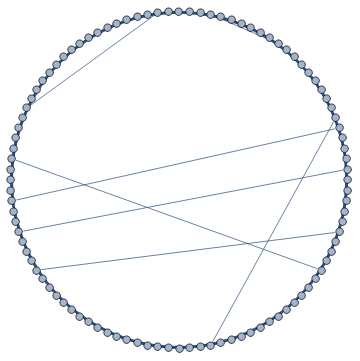}
 \includegraphics[width=2.7cm]{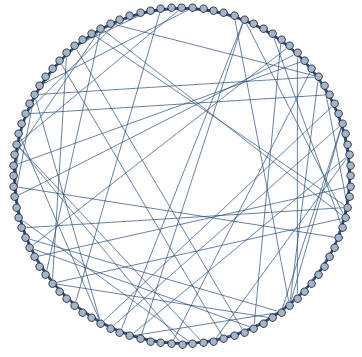}
  \caption{Example of a Watts--Strogatz rewiring algorithm for $n=100$-vertex graphs and rewiring probability $p=0,$ .01 and $0.1$ starting from a $2n$-regular graph. The larger $p$ the closer to a random graph (rewiring $p=1$).}
  \label{swplot}
\end{figure}

Observations of these properties have motivated the development of many models to explain these features. Of these, The Barab\'asi-Albert model~\cite{albert2002statistical} reproduces complex network features using a preferential attachment mechanism, 
and the Watts-Strogatz model~\cite{watts} also provides a mechanism for constructing small-world networks with a rewiring algorithm (which involves starting from a regular network and randomly rewiring); see for example Fig.~\ref{swplot}. 


\subsection{Network connectedness and complexity}

\begin{figure}[htbp!]
  \centering
\includegraphics[width=7.2cm]{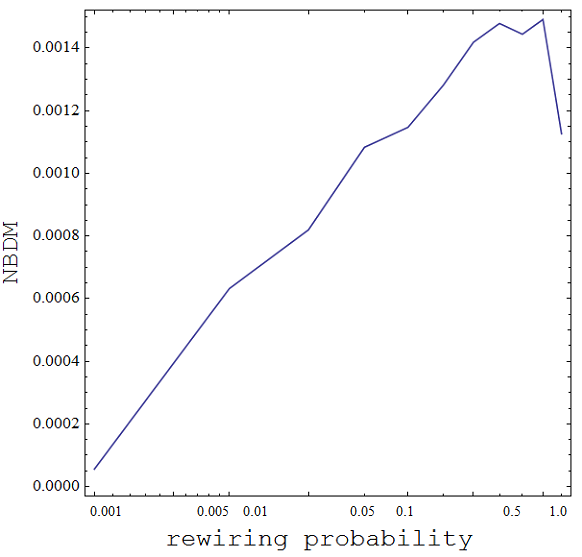}
  \caption{Kolmogorov complexity of the Watts-Strogatz model as a function of the rewiring probability on a 1000-node network starting from a regular ring lattice. Both the number of nodes and the number of links are kept constant, while $p$ varies; Kolmogorov complexity increases with $p$.}
  \label{connectednessplot}
\end{figure}

We have theoretically substantiated and experimentally demonstrated how network size, both in terms of the number of nodes and the density of edges for a fixed number of nodes, can impact Kolmogorov complexity values (above). However, Fig.~\ref{connectednessplot} demonstrates that node and edge count do not exclusively dominate $K$, as the graphs in the plot all have exactly the same graph size and edge density. The plot considers Watts--Strogatz graphs of size $V(g)=1000$ with rewiring probability ranging from $p=0$ (a regular graph) to $p=1$ (a random graph). The plot shows that the graph complexity increases with $p$, thus illustrating a change not subject to graph size or edge density, which are both the same for all cases. Rather, the increasing complexity must be due to other topological properties such as connectedness, link distribution and graph diameter.

\begin{figure}[htbp!]
  \centering
\includegraphics[width=6.7cm]{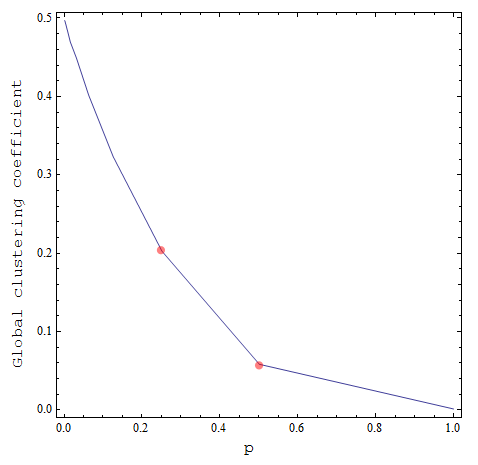}
\includegraphics[width=6.7cm]{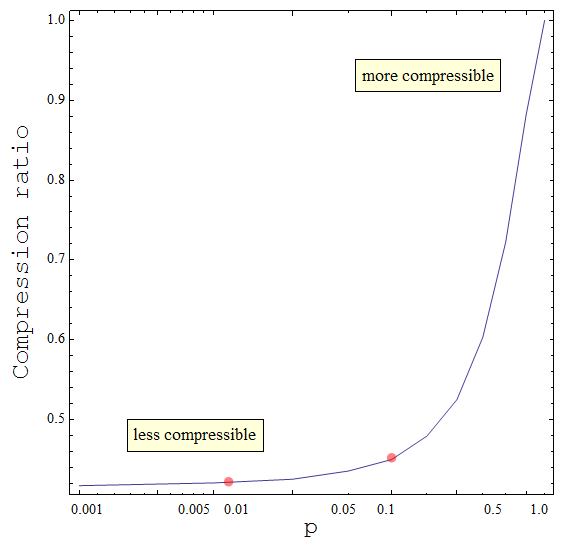}
  \caption{The Watts-Strogatz model starts from a ring lattice, hence highly compressible, then with rewiring probability $p$ ($x$-axis) the global clustering coefficient drops fast (Top) in this $15\times10^3$-node W-S network while approaching a random graph ($p=1$) slowing down at points $p \sim 0.2$ and 0.5  (red dots in plots) where compression ratios (Bottom) also display slight slope variations.}
  \label{clustering}
\end{figure}

\subsection{Topological characterization of artificial complex networks}

\begin{figure}[htbp!]
  \centering
  \includegraphics[width=8.5cm]{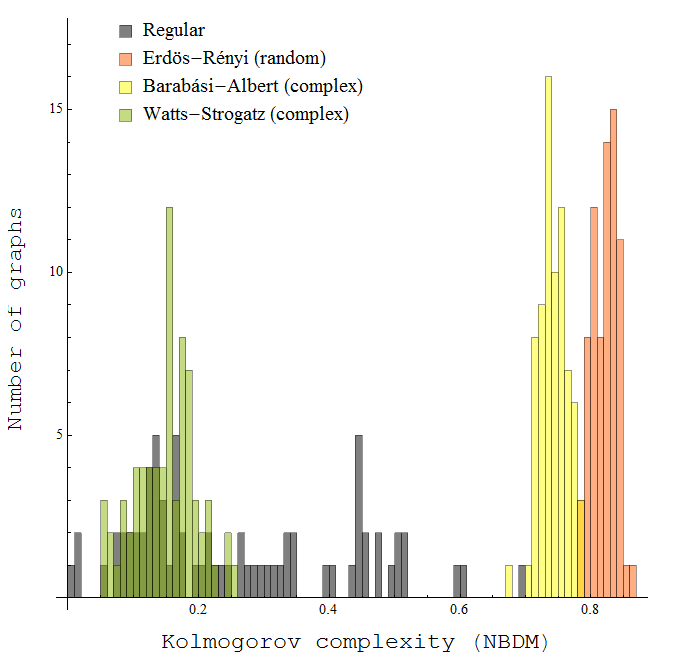}
  \caption{Distribution of 292 regular, Watts-Strogatz, Barab\'asi-Albert and Erd\"os-R\'enyi networks with $V(g)=30$ (73 networks each) with W-S rewiring probability $p=0.05$. The number 73 comes from the number of regular graphs of size $V(g)=30$ in the used repository (\emph{Mathematica}'s GraphData[]).}
  \label{fig10b}
\end{figure}

\begin{figure}[htbp!]
  \centering
  \includegraphics[width=8.5cm]{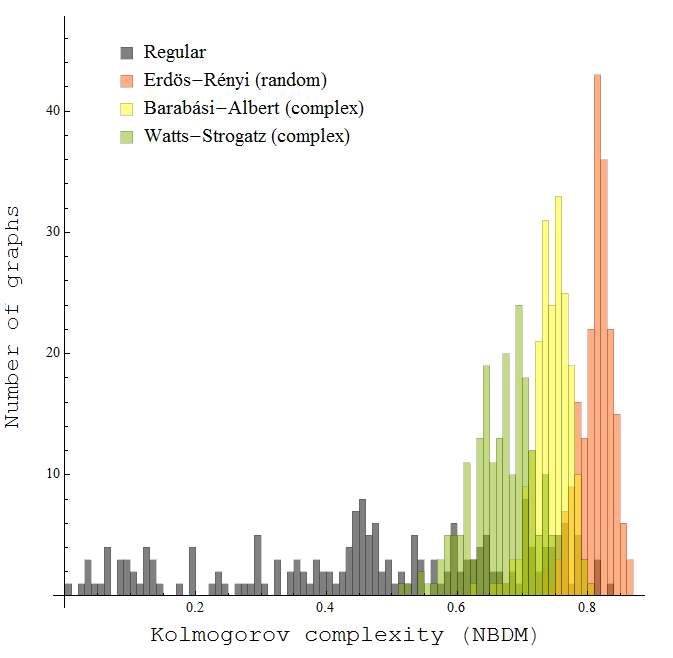}
  \caption{Distribution of 792 regular, Watts-Strogatz, Barab\'asi-Albert and Erd\"os-R\'enyi networks with $V(g)=20$ (198 networks each) with W-S rewiring probability from $p=0.05$ in Fig.~\ref{fig10b} to $p = 0.5$ as an experiment introducing randomness to witness the shift of the W-S networks towards higher complexity values of random graphs. For combinatorial reasons, there are more regular networks of size 30 than 20, hence this time the number 198 comes from the number of regular graphs found in the (\emph{Mathematica}'s GraphData[]).}
  \label{fig10b2}
\end{figure}

Random networks and Barab\'asi-Albert networks not only have exactly the same vertex count $V(g)$ in the experiment the results of which are summarised in Fig.~\ref{fig10b} and~\ref{fig10b2}, but also the same number of edges on average. The $K_m$ difference can only therefore be attributed to other topological properties related to each network model. Observing that $K$ can be affected by topological features such as clustering coefficient (as shown in~\ref{clustering} for Watts-Strogatz networks), we proceeded to examine  other network models.   
As shown in Figs.~\ref{fig10b} and~\ref{fig10b2}, $K_m$ approximated by the BDM assigns low Kolmogorov complexity to regular graphs and Watts-Strogatz networks and higher complexity to Barab\'asi-Albert networks, with random networks displaying the greatest Kolmogorov complexity as expected. Indeed, that random graphs are the most algorithmically complex is clear from a theoretical point of view: nearly all long binary strings are algorithmically random, and so nearly all random unlabelled graphs are algorithmically random~\cite{li}.



Barab\'asi-Albert networks are often referred to as scale free, because the node distribution follows a power law. However for small graphs such as those we analyse, it is questionable whether the global scaling is meaningful. An advantage of the method is, however, that it still differentiates between different network models despite their  small size. The theory of algorithmic information formally characterizes any object in terms of the properties specified in its description from which the object can be fully recovered. It is therefore vis-\`a-vis small objects that the theory presents its greatest challenges, given that the invariance theorem (see~\cite{li}) does not tell us anything about the rate of convergence in values.

We also considered regular graphs for this comparison including Haars, circulants, noncayley transitives, snarks, cubics, books, lattices and suns among other types of regular networks. The average NBDM complexity value of the regular graph group remained very low (and the distribution mostly uniform with a slight peak at around 0.5). The group of complex networks peak at different $K_m$ values, with Watts-Strogatz networks ranking low for $p=0.01$ when the small world effect is produced, and then moves towards high Kolmogorov complexity when rewiring probability $p$ increases. Barab\'asi-Albert networks peak at NBDM value equal to 0.75 and random graphs (Erd\"os-R\'enyi ) ranked the highest at an estimated Kolmogorov complexity value close to 0.9.

\section{Concluding remarks}

Our investigation connected disparate seminal subjects in graph theory and complexity science. We have shown computationally that Kolmogorov complexity approximations capture important group-theoretic and topological properties of graphs and networks, properties related to symmetry, density and connectedness, to mention a few and the most important. For example, we found that graphs with a large number of non-trivial automorphism groups tend to have smaller Kolmogorov complexity values. This was verified both for small artificial graphs and larger (but still rather small) empirical networks ranging from biological to social networks. 

We have also shown that graphs with differing algorithmic randomness distinguish models of networks, two of which are complex networks with different edge generation mechanisms. A number of connections between graph and network properties to complexity that other papers have claimed or have tried to connect before with hand-tailored measures were obtained naturally with Kolmogorov complexity. 

In future work, it may be interesting to take one class of networks of a fixed size and analyse what aspects of the topology change $K$, and what physical interpretations and implications may be associated to low and high complexities. Additionally, due to the small graphs we have analysed, it would be interesting to explore how approximations to $K$ behave for increasing network size for each model, and also with larger samples of graphs. Finally, extending these results to directed graphs would be another possible direction to explore. The findings suggest that analysing natural complex systems via Kolmogorov complexity may open new interesting avenues of research.

Supplemental material, including all the source code in \emph{Mathematica} v.9 can be downloaded (and read with the free \emph{Mathematica} player) at \url{http://complexitycalculator.org/graphcomplexitysuplementalmaterialv1.zip}.

\section*{Acknowledgements}

We thank Sebastian Ahnert, Jesper Tegn\'er and the anonymous referees for the helpful discussion and comments on the manuscript.

\bibliographystyle{elsarticle-num}

\bibliography{KolmoBib}

\pagebreak

\begin{widetext}
\newpage

\section*{Appendix}

This Appendix contains some additional figures to the main text.

\begin{figure*}[htbp!]
  \centering
  \includegraphics[width=16.1cm]{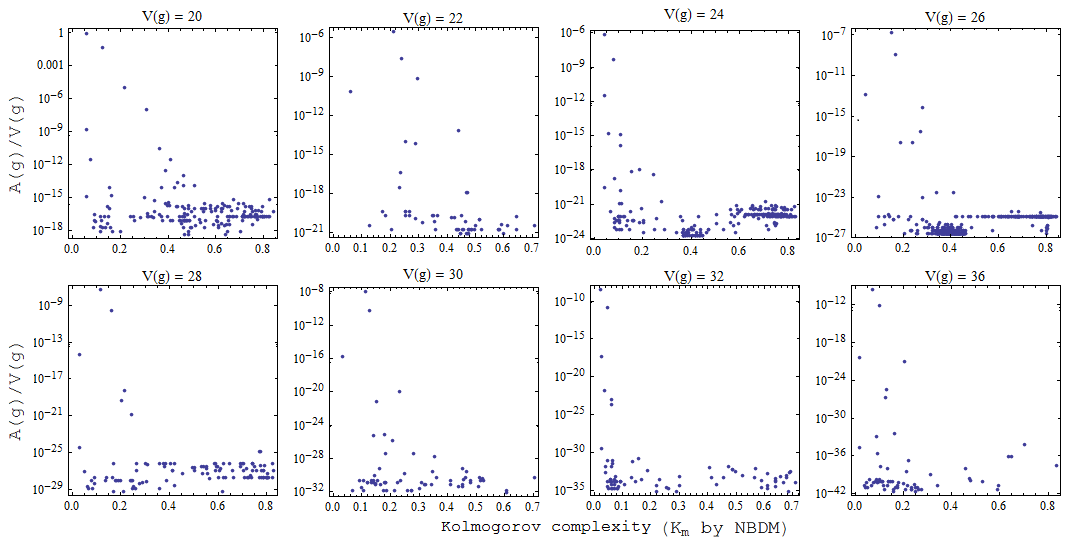}
  \caption{\label{fig8} Plots of number of graph automorphisms normalised by maximum number of edges of $g$, $A(g)/V(g)!$ ($y$-axis) versus (normalised) Kolmogorov complexity ($x$-axis) estimated by NBDM for connected regular graphs found in \emph{Mathematica} (GraphData[]) with size $V(g)=20$ to 36 nodes (only vertex sizes for which at least 20 graphs were found in the dataset were plotted). The decay can be seen, though the relationship is noisy.}
\end{figure*}

\begin{figure*}[htbp!]
  \centering
  \includegraphics[width=16.1cm]{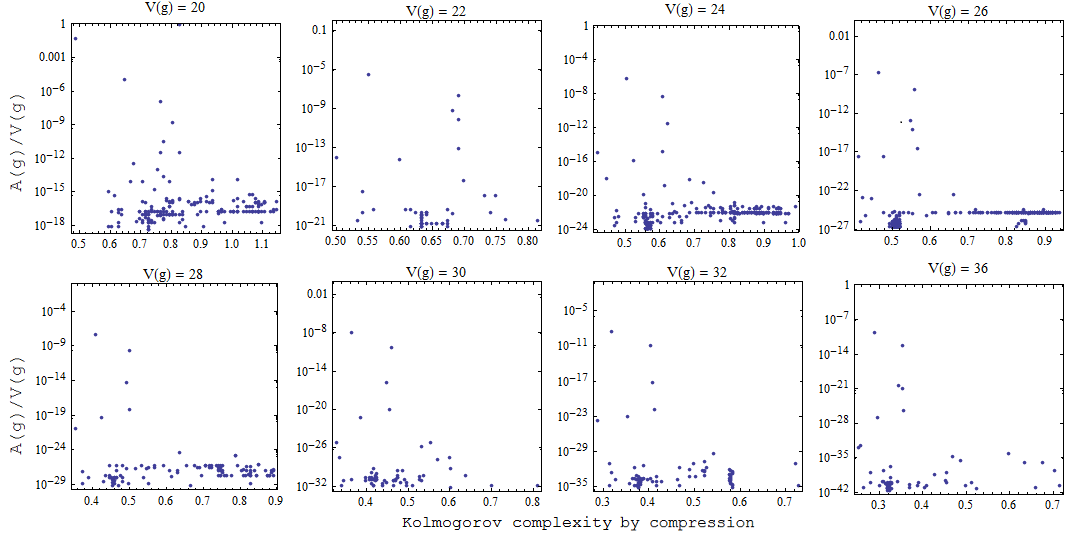}
  \caption{\label{fig9} Plots of number of graph automorphisms normalised by maximum number of edges of $g$, $A(g)/V(g)!$ ($y$-axis) versus Kolmogorov complexity ($x$-axis) estimated by lossless compressed length (Deflate) of connected regular graphs in \emph{Mathematica}'s GraphData[] with size $V(g)=20$ to 36 nodes (only vertex sizes for which at least 20 graphs were found in the dataset were plotted).}
\end{figure*}

\newpage

\begin{table}\centering
\ra{1.7}
\begin{tabular}{@{}l|c|c|r@{}}\hline
&  & \multicolumn{1}{c|}{\textbf{Normalised}} & \\ 
\multicolumn{1}{c|}{\textbf{Network description} (\textbf{$g$})}& \multicolumn{1}{c|}{$V(g)$} & \multicolumn{1}{c|}{$K_m(g)$ (BDM)} &  \multicolumn{1}{c}{$A(g)/V(g)$}\\ \hline \hline
Metabolic Network of Actinobacillus &&&\\ 
Actinomycetemcomitans & 993&0.00336&$4.42\times 10^{74}$ \\ 
Metabolic Network Neisseria Meningitidis&981&0.00344&$2.86\times 10^{76}$ \\ 
Perl Module Authors Network&840&0.00350&$4.63\times 10^{470}$ \\ 
Metabolic Network Campylobacter Jejuni&946&0.00370&$6.97\times 10^{74}$ \\ 
Metabolic Network Emericella Nidulans&916&0.00378&$3.43\times 10^{68}$ \\ 
Pyrococcus Horikoshii Network &953&0.00382&$4.22\times 10^{70}$ \\ 
Pyrococcus Furiosus Network&931&0.00384&$3.37\times 10^{68}$ \\ 
Metabolic Network Thermotoga Maritima&830&0.00477&$2.05\times 10^{64}$ \\ 
Mycoplasma Genitalium Network&878&0.00480&$6.75\times 10^{92}$ \\ 
Treponema Pallidum Network&899&0.00499&$2.71\times 10^{84}$ \\ 
Chlamydia Trachomatis Network&822&0.00511&$1.73\times 10^{75}$ \\ 
Metabolic Network Pyrococcus Furiosus&751&0.00511&$2.86\times 10^{50}$ \\ 
Rickettsia Prowazekii Network&817&0.00523&$1.39\times 10^{76}$ \\ 
Arabidopsis Thaliana Network&768&0.00535&$1.93\times 10^{60}$ \\ 
Oryza Sativa Network&744&0.00569&$3.45\times 10^{57}$ \\ 
Chlamydia Pneumoniae Network&744&0.00635&$2.00\times 10^{70}$ \\ 
Metabolic Network Oryza Sativa&665&0.00640&$9.49\times 10^{47}$ \\ 
Metabolic Network Rickettsia Prowazekii&456&0.01080&$1.10\times 10^{34}$ \\ 
Metabolic Network Mycoplasma Pneumoniae&411&0.01280&$1.85\times 10^{28}$ \\ 
Metabolic Network Borrelia Burgdorferi&409&0.01460&$2.10\times 10^{36}$ \\ 
\hline
\end{tabular}
\caption{\label{table1} Random sample of 20 real-world networks~\cite{jeong,johnson} from the 88 included in the study (and plotted in Fig.~\ref{largenetworks}), sorted from smallest to largest estimated Kolmogorov complexity values (NBDM). While the (negative) correlation between $K_m$ and $V(g)$ is almost perfect (Pearson coefficient -0.95) the (negative) correlation between $K_m$ and $A(g)$ is significant (Pearson coefficient -0.178) after normalisation by $V(g)$ for the 88 elements. The full descriptions and sources of the networks are available as supplemental material at \url{http://www.complexitycalculator.com/graphcomplexitysuplementalmaterialv1.zip}.}
\end{table}

The correlation of $K_m$ and $V(g)$ is explained by the theory. $K$ (not $K_m$ but the true uncomputable value $K$) is strongly positively correlated to $V(g)$ because larger networks can always potentially reach higher complexity values compared to small networks.

\begin{figure*}[htbp!]
  \centering
  \includegraphics[width=17cm]{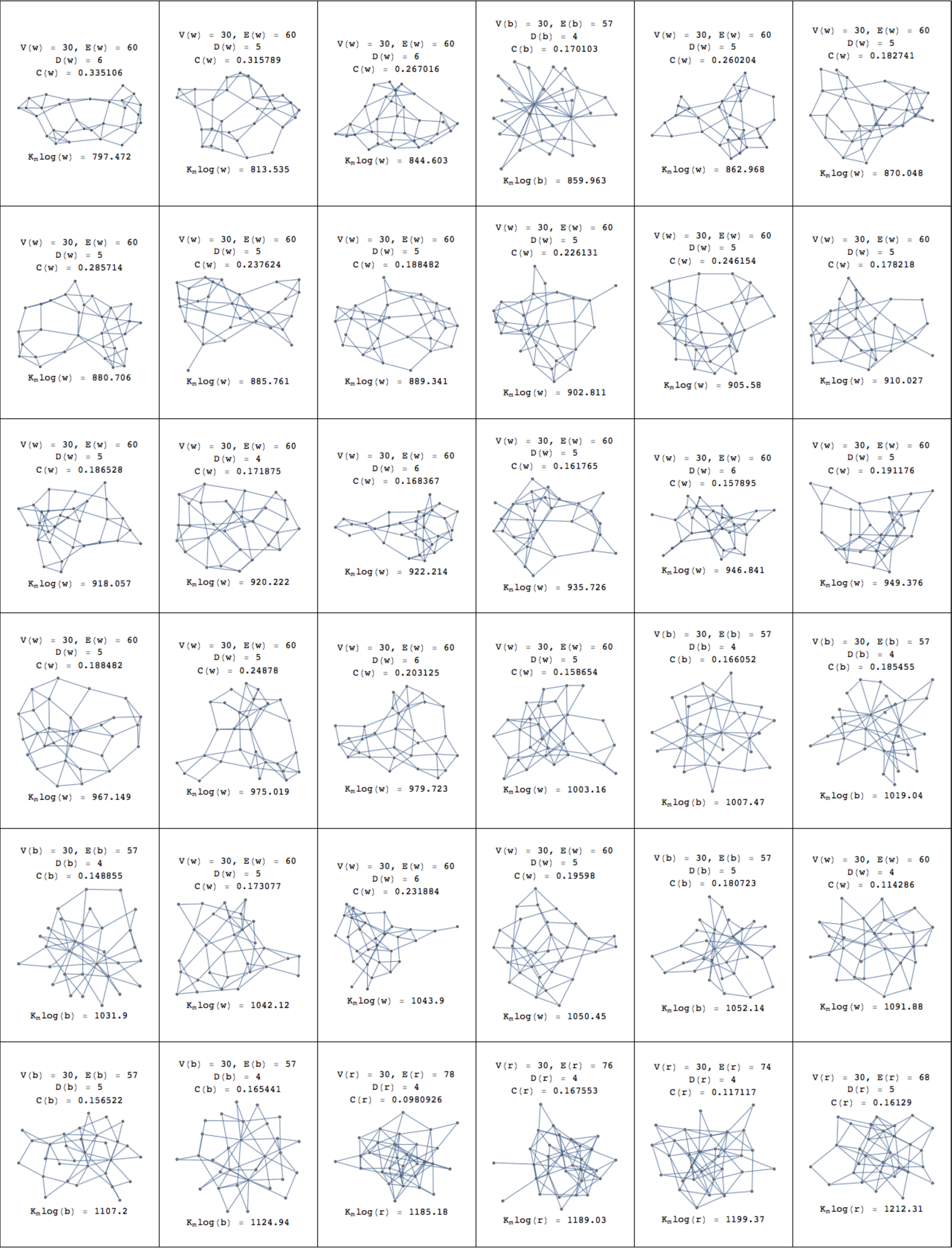}
  \caption{Random (Erd\"os--R\'enyi) graphs (denoted by $r$) versus complex networks (Watts-Strogatz and Barab\'asi-Albert) (denoted by $w$ and $b$) sorted by Kolmogorov complexity (smallest to largest $K_m\log$) as approximated by the BDM.}
  \label{fig11}
\end{figure*}

\end{widetext}

\end{document}